\documentclass[aps,prd,showpacs,nofootinbib,preprintnumbers]{revtex4}
\usepackage{amsmath}
\usepackage{amssymb}
\usepackage{amsfonts}
\usepackage{graphicx}
\usepackage{bm}

\begin{document}

\tolerance=5000

\title{On isotropic turbulence in the dark fluid universe}

\author{Iver Brevik$^1$, Olesya Gorbunova$^2$, Shin'ichi Nojiri$^{3,4}$, Sergei D. Odintsov$^5$}

\affiliation{
\ \\
${}^1$Department of Energy and Process Engineering, Norwegian University
of Science and Technology, N-7491 Trondheim, Norway~\\
${}^2$Dipartimento di Fisica, Universita di Trento
and Istituto Nazionale di Fisica Nucleare
Gruppo Collegato di Trento, Italia. Also at TSPU, Tomsk, Russia~\\
${}^3$Department of Physics, Nagoya University,~Nagoya 464-8602, Japan~\\
${}^4$Kobayashi-Maskawa Institute for the Origin of Particles and the
Universe, Nagoya University, Nagoya 464-8602, Japan~\\
${}^5$Institucio Catalana de Recerca i Estudis Avan~ACcats (ICREA)
and Institut de Ciencies de l'Espai (IEEC-CSIC), Campus UAB,
Facultat de Ciencies, Torre C5-Par-2a pl, E-08193 Bellaterra
(Barcelona), Spain. Also at TSPU, Tomsk, Russia }

\date{\today}

\begin{abstract}

As first part of this work, experimental information about the
decay of isotropic turbulence in ordinary hydrodynamics, $\overline{{\bf u}^2(t)}
\propto t^{-6/5}$, is used as input in FRW equations in order to
investigate how an initial fraction $f$ of turbulent kinetic
energy in the cosmic fluid influences the cosmological development
in the late, quintessence/phantom, universe. First order
perturbative theory to the first order in $f$ is employed. It
turns out that both in the Hubble factor, and in the energy
density, the influence from the turbulence fades away at late
times. The divergences in these quantities near the Big Rip behave
essentially as in a non-turbulent fluid. However, for the scale
factor, the turbulence modification turns out to diverge
logarithmically. As second part of our work, we consider the full
FRW equation in which the turbulent part of the dark energy is
accounted for by a separate term. It is demonstrated that
turbulence occurrence may change the future universe evolution due
to dissipation of dark energy. For instance, phantom-dominated
universe becomes asymptotically a de Sitter one in the future,
thus avoiding the Big Rip singularity.

\end{abstract}


\maketitle

\section{ Introduction}

Consider a spatially flat FRW universe with $H=\dot{a}/a$ the
Hubble parameter. In standard notation the FRW equation
\begin{equation}
\frac{\ddot{a}}{a}=-\frac{1}{6}\kappa^2(\rho+3p) \label{1}
\end{equation}
with $\kappa^2=8\pi G$, implies that for the scale factor $a(t)$
to depict a curve concave upwards when drawn as a function of $t$,
i.e. $\ddot{a}>0$, it is sufficient that the equation of state
(EoS) satisfies the condition
\begin{equation}
p<-\frac{1}{3}\rho\, . \label{2}
\end{equation}
In order to obtain $\dot{H}>0$ it follows, however, from the
equation
\begin{equation}
\dot{H}=-\frac{1}{2}\kappa^2(\rho+p) \label{3}
\end{equation}
that $p$ has to satisfy the stronger condition
\begin{equation}
p<-\rho\, . \label{4}
\end{equation}
This is the phantom region, corresponding to a  positive tensile stress in
the dark energy fluid.
It is known that phantom-dominated universe usually enters to finite-time
future singularity (Big Rip)
(see Refs.~\cite{Caldwell:2003vq,ref5,Nojiri:2005sx}).
The region $-\rho<p<-\rho/3$ is called the quintessence region, where the expansion of the
universe is accelerated $\ddot a>0$ but not super-accelerated $\dot H>0$.
Note that effective quintessence universe may also end up in one of three
possible types of future singularity \cite{Nojiri:2005sx}.

This paper analyzes the possibility whether there can be  a {\it turbulent microstructure}
superimposed on the dark fluid. Quite obviously it is  the isotropic version of turbulence theory
which then becomes  most relevant, as one wishes not to disturb the macroscopic isotropy.
Turbulence  generally implies that there is
a loss of kinetic energy into heat. Whereas this loss is often
described macroscopically, in terms of a bulk viscosity $\zeta$
(cf., for instance, Refs.~\cite{weinberg71,gron90,brevik94}), our
program here is to replace $\zeta$ with a microscopic {\it shear}
viscosity. In other words, we wish to replace the length scale
associated with the macroscopic $\zeta$ by the Kolmogorov
microlength  scale, conventionally denoted by $\eta$.

The next section gives a brief overview of isotropic classical
turbulence theory, in order to put our present approach in a
proper context. Thereafter, we consider how the turbulence effect
can be taken into account in the cosmological formalism. We focus
on the following two possibilities:

\noindent
1) The turbulence effect can be included by the addition of a constant {\it
fraction}, called  $f$, to the laminar ordinary energy density
$\rho$ in the first FRW equation. Assuming $f$ to be a small
quantity, a perturbative solution to the first order in $f$ can
conveniently be found.

\noindent
2) Our second approach is to write the total energy density as a
sum of four different parts: (i) a laminar dark energy part, (ii)
a turbulent dark energy part, (iii) a radiation part, and (iv) an
ordinary matter part. The fate of the universe is in principle
predictable, on the basis of the weight given to each of the
constituents of the total energy density.

Options 1) and 2) are considered in Sections \ref{frac} and \ref{turb},
respectively.

We mention finally that our method of decomposing the fluid into  turbulent and  non-turbulent parts is in principle similar to the method recently used by Balakin and Bochkarev \cite{balakin11}. These authors divided the energy as well as the pressure of the cosmic fluid into two components, one component referring to dark matter, the other referring to dark energy.

\section{Extracts from Kolmogorov's isotropic turbulence theory \cite{panchev71,landau87}}

Turbulence  generally implies
a loss of kinetic energy into heat. As mentioned, we replace  the macroscopic bulk viscosity
with a microscopic shear viscosity, corresponding to  the Kolmogorov length
\begin{equation}
\eta= \left( \frac{\nu^3}{\epsilon}\right)^{1/4}\, . \label{5}
\end{equation}
Here $\nu$ is the kinematic microscopic shear viscosity and $\epsilon$
is the dissipation per unit time and unit mass.

Let $l$ denote the external scale of the turbulence, with $1/l$ the corresponding wave number.
The large eddies move around with only a little dissipation of energy. According to the so-called
second hypothesis of Kolmogorov \cite{panchev71}, in an isotropic region the motion is entirely
determined by friction and inertia. There occurs a continuous flux of energy transferred by means
of a hierarchy of eddies corresponding to the dissipation $\epsilon$. Let $\lambda$ characterize
the size of an eddy, $k=1/\lambda$ being the corresponding wave number.
The {\it equilibrium range} is that for which all memory of the flow is lost,
\begin{equation}
k \gg 1/l\, . \label{6}
\end{equation}
If $u_{\lambda}$ is the typical velocity of an eddy of size $\lambda$, the internal Reynolds
number is Re$_{\lambda} \sim \lambda u_{\lambda}/\nu$.
For increasing values of $k$, Re$_{\lambda}$ decreases. Dissipation becomes important
when Re$_{\lambda}\sim 1$. This is just the condition leading to the Kolmogorov length (\ref{5}).

If the Reynolds number of the flow as a whole is high, the wave numbers $1/l$ and $1/\eta$
are widely separated, and there exists an inertial subrange characterized by
\begin{equation}
\frac{1}{l} \ll k \ll \frac{1}{\eta}, \label{7}
\end{equation}
in which the fluid behaves like a non-viscous fluid. A famous formula for the spectral energy density,
conventionally called $E(k)$, in the inertial subrange is
\begin{equation}
E(k)=\alpha \epsilon^{2/3}\,k^{-5/3}\, , \label{8}
\end {equation}
where $\alpha \approx 1.5$ is the Kolmogorov constant.

For practical purposes the von K\'{a}rm\'{a}n interpolation formula for $E(k)$, linking the region
of small $k$ to the region of high $k$, is useful in order
to calculate the total energy density $E$ by integrating over all wave numbers
(cf., for instance, Ref.~\cite{brevik92}). We abstain from going into further
detail here. Of interest for us here is, however, is the empirical decay law for isotropic
turbulence (of course, under the assumption that it is left to itself; there are no external sources).
Based on grid experiments in wind and water tunnels, it turns out that the mean kinetic energy $\frac{1}{2}\overline {{\bf{u}}^2(t)}$ decays as
\begin{equation}
 \overline {{\bf{u}}^2(t)}  \propto t^{-6/5}; \label{9}
\end{equation}
cf. Refs.~\cite{sreenivasan80,rosen87}, as well as the theoretical treatment in \cite{brevik92}.
We shall make use of this relationship in the following.

Consider now the classical equation of motion for a viscous fluid:
\begin{equation}
\partial_t (\rho_mu_i)+\partial_k\Pi_{ik}=0, \label{A}
\end{equation}
where $\rho_m$ is the mass density and $\Pi_{ik}$ the momentum
flux density tensor \cite{landau87},
\begin{equation}
\Pi_{ik}=p\delta_{ik}+\rho_mu_iu_k-\mu
(\partial_ku_i+\partial_iu_k), \label{B}
\end{equation}
$\mu$ being the shear viscosity. Taking the mean of this equation,
observing that $\bar{u}_i=0$ in homogeneous and isotropic
turbulence, we get
\begin{equation}
\bar{\Pi}_{xx}=\bar{\Pi}_{yy}=\bar{\Pi}_{zz} \equiv p_{\rm
eff}=p+\frac{2}{3}\rho_{\rm turb}, \label{C}
\end{equation}
where $p_{\rm eff}$ is the effective which takes into account that the
thermodynamical pressure is augmented by a term
$(2/3)\rho_{\rm turb}$ associated with the turbulent energy
density,
\begin{equation}
\rho_{\rm turb}=\frac{1}{2}\rho_m {\overline{u_i u_i}} \equiv
\frac{1}{2}\rho_m \overline{{\bf{u}}^2}. \label{D}
\end{equation}
Thus $\rho_{\rm turb}$ designates a mean quantity.

\section{FRW equations with a fraction $f$ of the energy as turbulent energy \label{frac}}

We will now consider this fluid in a cosmological setting. As
usual in turbulence theory we may start by decomposing the fluid
velocity $u_i$ into a mean component $U_i$ and a fluctuating
component $u_i'$, $u_i=U_i+u_i'$. However, in a comoving reference
frame $U_i=0$, so that we can simply replace $u_i'$ with $u_i$. In
order to keep the turbulent part of the cosmic fluid separate from
the non-viscous (non-turbulent) part, we shall from now on endow
non-viscous quantities with a subscript zero. Thus Eq.~(\ref{C})
is rewritten as
\begin{equation}
p_{\rm eff}=p_0+\frac{2}{3}\rho_{\rm turb}, \label{E}
\end{equation}
with $ \rho_{\rm turb}= \frac{1}{2}\rho_m \overline{{\bf{u}}^2}$
as before. Note that the turbulent energy in the comoving frame is
regarded as a nonrelativistic  quantity. The total energy
density can be written as
\begin{equation}
\rho=\rho_0+\rho_{\rm turb}. \label{F}
\end{equation}
As mentioned above, we shall first look for a solution of the
cosmological equations when a constant fraction $f$ of the energy
exists in the form of turbulent energy. As we are primarily
interested in the dark energy epoch of the universe, we assume
henceforth that the thermodynamical parameter $w$ satisfies the
inequality $w<-1$ (see Eq.~(\ref{L}) below). The initial instant
for our considerations will be denoted by $t_{\rm in}$.

By assumption we can thus write the energy density at time
$t=t_{\rm in}$ as
\begin{equation}
\rho(t_{\rm in})=\rho_0(t_{\rm in})(1+f). \label{G}
\end{equation}
We see that $f$ can be interpreted as the ratio between the turbulent
energy and the total energy  at $t_{\rm in}$,
\begin{equation}
f=\frac{\rho_{\rm turb}}{\rho_0} \Big|_{t_{\rm in}}. \label{H}
\end{equation}
As $\rho_0$ includes the rest mass, this shows showing that the assumption $f \ll 1$ is a plausible one.

For $t>t_{\rm in}$ we now require $\rho_{\rm turb}$ to decay with
time as
\begin{equation}
\rho_{\rm turb} \propto t^{-6/5}, \label{I}
\end{equation}
in accordance with Eq.~(\ref{9}) for ordinary turbulence. We can
write $\rho(t)$ for $t \geq t_{\rm in}$ in the form
\begin{equation}
\rho(t)=\rho_0(t)[1+f\rho_1(t)], \label{J}
\end{equation}
with
\begin{equation}
\rho_1(t)=\left(\frac{t_{\rm in}}{t}\right)^{6/5}. \label{K}
\end{equation}
When $t=t_{\rm in}$, this agrees with Eq.~(\ref{G}).

The equation of state (EoS) for the cosmic fluid is now to be
introduced. This can be done in various ways. We choose write it
in such a way that only non-turbulent quantities are involved,
\begin{equation}
p_0(t)=w\rho_0(t), \quad w:~\rm constant. \label{L}
\end{equation}
The effective pressure will analogously to Eq.~(\ref{J}) be
expanded as
\begin{equation}
p_{\rm eff}(t)=w\rho_0(t)[1+fp_1(t)]. \label{M}
\end{equation}
At $t=t_{\rm in}$ it  follows that $p_1(t_{\rm in})$ is
actually a known quantity,
\begin{equation}
wp_1(t_{\rm in})=\frac{2}{3}. \label{N}
\end{equation}

In the same way we can expand the scale factor as
\begin{equation}
a(t)=a_0(t)[1+fa_1(t)]\, , \label{16}
\end{equation}
and analogously for the Hubble factor
\begin{equation}
H(t)=H_0(t)[1+f H_1(t)]. \label{17}
\end{equation}
The correction terms $\{p_1,a_1,H_1\}$ are all of zeroth order in $f$.

From $H=\dot{a}/a$ we get at once
\begin{equation}
\dot{a}_1=H_0H_1, \label{18}
\end{equation}
whereas the first FRW equation  $H^2=\frac{1}{3}\kappa^2\rho
$ yields the first order relationship
\begin{equation}
H_1=\frac{1}{2}\rho_1. \label{19}
\end{equation}
The Hubble parameter $H(t)$ thus satisfies the equation
\begin{equation}
H(t)=H_0(t)\left[ 1
+\frac{1}{2}f\left(\frac{t_\mathrm{in}}{t}\right)^{6/5}\right]\, . \label{20}
\end{equation}
We still need to determine $H_0(t)$. It can be found from the
non-turbulent FRW equations
\begin{equation}
H_0^2=\frac{1}{3}\kappa^2 \rho_0\, , \label{21}
\end{equation}
\begin{equation}
\frac{\ddot{a}_0}{a_0}+\frac{1}{2}H_0^2=-\frac{1}{2}\kappa^2p_0\, ,
\label{22}
\end{equation}
which, together with the conservation equation for energy,
${T^{0\nu}}_{;\nu}=0$, yield
\begin{equation}
\dot{\rho}_0+3H_0(\rho_0+p_0)=0\, . \label{23}
\end{equation}
  From Eq.~(\ref{21}), $\dot{H}_0=({\sqrt{3}}/6)\kappa \dot{\rho}_0/{\sqrt{\rho}_0}$,
and as $\dot{H}=-H^2+\ddot{a}/a$ we get
\begin{equation}
\dot{H}_0+\frac{3}{2}\gamma H_0^2=0\, . \label{24}
\end{equation}
Here we have for convenience introduced the symbol $\gamma$,
defined as
\begin{equation}
\gamma=1+w. \label{25}
\end{equation}
The solution of this equation is
\begin{equation}
H_0(t)=\frac{H_0(t_\mathrm{in})}{1
+\frac{3}{2}\gamma H_0(t_\mathrm{in})(t-t_\mathrm{in})}\, . \label{26}
\end{equation}
Thus
\begin{equation}
H(t)=\frac{H_0(t_\mathrm{in})}{1
+\frac{3}{2}\gamma H_0(t_\mathrm{in})(t-t_\mathrm{in})}
\left[ 1+\frac{1}{2}f\left(\frac{t_\mathrm{in}}{t}\right)^{6/5}\right]\, . \label{27}
\end{equation}
Correspondingly, we obtain
\begin{equation}
\rho(t)=\frac{3}{\kappa^2}
\frac{H_0^2(t_\mathrm{in})}{[1+\frac{3}{2}\gamma H_0(t_\mathrm{in})(t-t_\mathrm{in})]^2}\left[
1+f\left(\frac{t_\mathrm{in}}{t}\right)^{6/5}\right]\, . \label{28}
\end{equation}
We can now draw the following important conclusion: both for the
Hubble factor, and the energy density, the influence from the
turbulence fades away when $t \gg t_\mathrm{in}$. To a good
approximation the future singularity  occurs at the same instant
$t=t_s $ as if turbulence were absent, i.e.
\begin{equation}
t_s=t_\mathrm{in} + \frac{2}{3|\gamma|H_0(t_\mathrm{in})}\, . \label{29}
\end{equation}
Both $H$ and $\rho$ diverge at $t=t_s$. Near $t_s$, as
$t_\mathrm{in}/t_s \ll 1$,
\begin{equation}
H(t) \approx \frac{H_0(t_\mathrm{in})}{1-t/t_s}, \quad t \rightarrow
t_s\, , \label{30}
\end{equation}
\begin{equation}
\rho(t) \approx \frac{3}{\kappa^2}
\frac{H_0(t_\mathrm{in})}{(t-t/t_s)^2}\, , \quad t\rightarrow t_s\, . \label{31}
\end{equation}
Consider next the correction $a_1$ to the scale factor. From
Eqs.~(\ref{18}) and (\ref{19}),
\begin{equation}
\dot{a}_1=\frac{1}{2}H_0\rho_1\, , \label{31B}
\end{equation}
from which we obtain by integration, setting $x=t/t_\mathrm{in}-1$,
\begin{equation}
a_1(t)=\frac{1}{2}H_0(t_\mathrm{in})t_\mathrm{in}\int_0^{t/t_\mathrm{in}-1}
\frac{dx}{(1+x)^{6/5}}\frac{1}{1-(t_\mathrm{in}/t_s)x}. \label{32}
\end{equation}
We need not calculate this integral in full, but note that it
diverges logarithmically at $x=t_s/t_\mathrm{in}$. Omitting
multiplicative factors, we write for the dominant part
\begin{equation}
a_1(t) \sim \ln(1-t/t_s), \quad t\rightarrow t_s\, . \label{33}
\end{equation}
Thus the modification coming from turbulence is in this case
itself turbulent. As the solution in the case $f=0$ is now
\begin{equation}
a_0(t)=\frac{a_0(t_\mathrm{in})} {[1+\frac{3}{2}\gamma
H_0(t_\mathrm{in})(t-t_\mathrm{in}) ]^{2/3|\gamma|}}\, , \label{34}
\end{equation}
it follows however from the expansion (\ref{16}) that the
divergence in $a_0(t)$ is much stronger than the turbulence
modification. The dominant term near $t=t_s$ is thus
\begin{equation}
a(t) \approx \frac{a_0(t_\mathrm{in})}{(1-t/t_s)^{2/3|\gamma|}}\, ,
\quad t\rightarrow t_s\, , \label{36}
\end{equation}
just as in the case $f=0$.

So far, we have made use of the first FRW equation only; we have not considered the pressure in the cosmic fluid. To deal with the pressure, we have to take into account the second FRW equation also, for instance in the form
\begin{equation}
\frac{d}{dt}(\rho a^3)=-3H p_{\rm eff}a^3. \label{37}
\end{equation}
By expanding in the parameter $f$ in the same way as above, we obtain to first order
\begin{equation}
p_1=\rho_1-H_1-\frac{\dot{\rho_1}+3\dot{a}_1}{3H_0 w}\, .
\label{38}
\end{equation}
Inserting for $\rho_1$, $H_1$, $a_1$, and $H_0$ we get
\begin{equation}
p_1(t)=\frac{1}{2}\left[
1-\frac{1}{w}+\frac{4}{5}\frac{1}{H_0wt}\right]\rho_1(t)\, .
\label{39}
\end{equation}
It is here to be observed that if one extrapolates this expression back in time, until
the initial instant $t_{\rm in}$, the expression does not in general agree with
the previous equation (\ref{N}). The reason for this is that our condition (\ref{N})
on the initial pressure makes the system mathematically over-determined.
Of most physical interest is, however, the cosmic pressure at late times,
$H_0$ and $t$ large, in which case the last term in Eq.~(\ref{39}) fades away and we get
\begin{equation}
p_1(t) \approx
\frac{1}{2}\left(1-\frac{1}{w}\right)\left(\frac{t_{in}}{t_s}\right)^{6/5}\, ,
\quad t\rightarrow t_s\, . \label{40}
\end{equation}
When $w$ lies between -1/3 and -1, i.e. in the
quintessence region, the value of $p_1$ is actually higher than
when $w<-1$.

\section{Turbulent dark energy density component as a separate term
in the FRW equation \label{turb}}

We now leave the perturbative approach, and consider instead the
first FRW equation together with the energy conservation
equation when the total energy density is written as a sum of four
different parts: first, a dark energy contribution consisting of a
laminar part $\rho_\mathrm{dark}$ and a turbulent energy part
$\rho_\mathrm{turb}$ so that
\begin{equation}
\rho_\mathrm{dark~energy}=\rho_\mathrm{dark}+\rho_\mathrm{turb}; \label{41}
\end{equation}
secondly, a radiation part $\rho_\mathrm{rad}$ and an ordinary matter
part $\rho_\mathrm{matter}$. The FRW equation thus reads
\begin{equation}
\frac{3}{\kappa^2}H^2=\rho_\mathrm{dark}+\rho_\mathrm{turb}
+\rho_\mathrm{rad}+\rho_\mathrm{matter}\, . \label{42}
\end{equation}
As in the previous section, we follow the development of the
universe from the instant $t_\mathrm{in}$ onwards, and we adopt the
same empirical law for the time development of the turbulent
energy density,
\begin{equation}
\rho_\mathrm{turb}=\rho_\mathrm{turb}(t_\mathrm{in})
\left(\frac{t}{t_\mathrm{in}}\right)^{-6/5}\, . \label{43}
\end{equation}
The time derivative will be written in the form
\begin{equation}
\dot{\rho}_\mathrm{turb}=-C\rho_\mathrm{turb}^{11/6}\, , \label{44}
\end{equation}
where
\begin{equation}
C=\frac{6}{5t_\mathrm{in}}\left[\rho_\mathrm{turb}
(t_\mathrm{in})\right]^{-5/6}\, . \label{45}
\end{equation}
Now consider the energy balance equation in which
$\dot{\rho}_\mathrm{turb}$ is considered as a source term,
\begin{equation}
\dot{\rho}_\mathrm{turb}+3H(\rho_\mathrm{turb}+p_\mathrm{turb})
=-C\rho_\mathrm{turb}^{11/6}\, . \label{46}
\end{equation}
Here, we assume Eq.~(\ref{44}) should hold only in the flat universe. In
 the FRW universe, Eq.~(\ref{44}) should be changed as in (\ref{46}).
Because of  the turbulence, the kinetic energy changes into heat and the
heat then becomes radiation. Then we may consider the the
conservation law for radiation in the form
\begin{equation}
\dot{\rho}_\mathrm{rad}+3H(\rho_\mathrm{rad}+p_\mathrm{rad})
=C\rho_\mathrm{turb}^{11/6}\, . \label{47}
\end{equation}
Here, $p_\mathrm{rad}=\rho_\mathrm{rad}/3$. For definiteness let us
consider the case where the turbulent part $\rho_\mathrm{turb}$
dominates. Then Eq.~(\ref{42}) gives
\begin{equation}
H \sim \frac{\kappa}{\sqrt 3}\,\rho_\mathrm{turb}^{1/2}\, . \label{48}
\end{equation}
We shall now assume that the EoS parameter $w_\mathrm{turb}$ of the
turbulent part is constant. The EoS for the turbulent quantities
is written in conventional form,
\begin{equation}
p_\mathrm{turb}=w_\mathrm{turb}\,\rho_\mathrm{turb}\, . \label{49}
\end{equation}
Then, with the definition $\gamma_\mathrm{turb}=1+w_\mathrm{turb}$ we
can write Eq.~(\ref{46}) as
\begin{equation}
0 = \dot{ \rho}_{\mathrm{turb}} +  \kappa \sqrt{3} \,\gamma_{\rm
turb}\, \rho_{\mathrm{turb}}^{3/2} + C
\rho_{\mathrm{turb}}^{11/6}\, . \label{47B}
\end{equation}
We shall discuss three different options for this equation:

\noindent
(i) If
\begin{equation}
\rho_\mathrm{turb} \gg \left( \frac{\kappa}{C}\right)^3\, , \label{48B}
\end{equation}
the third term dominates compared with the second, and we recover
the expression (\ref{44}). It means that $\rho_\mathrm{turb}$ behaves
as in flat spacetime, $\rho_\mathrm{turb} \propto t^{-6/5}$.

\noindent
(ii) By contrast, if
\begin{equation}
\rho_\mathrm{turb} \ll \left( \frac{\kappa}{C}\right)^3\, ,\label{49B}
\end{equation}
the second term dominates compared with the third, and the
turbulent term becomes negligible. Then $\rho_\mathrm{turb}$ behaves
as a usual perfect fluid, giving $\rho_\mathrm{turb} \propto
a^{-3\gamma_\mathrm{turb}}$ and
$H \sim \frac{2}{3\gamma_\mathrm{turb}}\frac{1}{t}$
(it is here assumed that $w_\mathrm{turb}>-1$).

\noindent
(iii) If
\begin{equation}
w_\mathrm{turb}<-1\, , \label{50}
\end{equation}
there exists remarkably enough a solution where $\rho_\mathrm{turb}$
is a constant,
\begin{equation}
\rho_\mathrm{turb}=\left(-\frac{C}{3\gamma_\mathrm{turb}}\right)^{-6/5}\, .
\label{51}
\end{equation}
We now consider the case where the scale factor $a$ is given as a
function of the cosmological time, $a=a(t)$. Equation (\ref{46})
may be rewritten as
\begin{equation}
\frac{d}{dt}\left( a^{3\gamma_\mathrm{turb}}\rho_\mathrm{turb}\right)=
  -C a^{-5\gamma_\mathrm{turb}/2}\left(a^{3\gamma_\mathrm{turb}}
\rho_\mathrm{turb}\right)^{11/6}\, , \label{52}
\end{equation}
which can be integrated to yield
\begin{equation}
\rho_\mathrm{turb}(t)
=a(t)^{-3\gamma_\mathrm{turb}}\left(\frac{5C}{6}\int_{t_\mathrm{in}}^tdt'
a(t')^{-5\gamma_\mathrm{turb}/2} + C_0\right)^{-6/5}\, . \label{53}
\end{equation}
Here $C_0$ is a constant of  integration, which can be determined
from the initial condition at $t=t_\mathrm{in}$. When $a(t)$ is a
constant, Eq.~(\ref{53}) reproduces the standard result:
$\rho_\mathrm{turb} \propto t^{-\frac{6}{5}}$ when $t$ is large
enough. In case of $w_\mathrm{turb} =-1$, even if $a(t)$ is not a
constant, we obtain $\rho_\mathrm{turb} \propto t^{-\frac{6}{5}}$
for large $t$, again.

We can also integrate (\ref{47}) to obtain
\begin{equation}
\label{TR1}
\rho_\mathrm{rad} (t) = C a(t)^{-4} \left(\int_{t_\mathrm{in}}^t dt' a(t')^4 \rho_\mathrm{turb} (t)
+ C_1 \right)\, .
\end{equation}
Here $C_1$ is a constant of  integration.

When $w_\mathrm{turb} \neq -1$, if we consider the case of the de Sitter space: $a(t) = a_0 \mathrm{e}^{H_0 t}$
with constants $a_0$ and $H_0$, we obtain
\begin{equation}
\label{T7} \rho_\mathrm{turb} = \mathrm{e}^{-3
\gamma_\mathrm{turb} H_0 t} \left( \frac{C}{3 \gamma_\mathrm{turb}
}\left( \mathrm{e}^{-\frac{5 \gamma_\mathrm{turb} H_0}{2}
t_\mathrm{in}}
 - \mathrm{e}^{-\frac{5\gamma_\mathrm{turb} H_0}{2} t} \right)
+ \frac{2C_0 a_0^{\frac{5 \gamma_\mathrm{turb} H_0}{2}}}{5
\gamma_\mathrm{turb} }\right)^{-\frac{6}{5}}\, .
\end{equation}
When $w>-1$ and $t$ is large enough, we find
\begin{equation}
\label{T8} \rho_\mathrm{turb} \propto \mathrm{e}^{-3 \gamma
_\mathrm{turb} H_0 t } \, .
\end{equation}
On the other hand, when $w<-1$ and $t\gg t_\mathrm{in}$,
$\rho_\mathrm{turb}$ goes to a constant
\begin{equation}
\label{T9} \rho_\mathrm{turb} \to \left( - \frac{C}{3 \gamma
_\mathrm{turb}  } \right)^{-\frac{6}{5}}\, ,
\end{equation}
which corresponds to (\ref{51}).

When $w_\mathrm{turb} \neq -1$, if we consider the case of an
effective quintessence-like power law expansion, $a(t)=a_0
t^{h_0}$ with constants $a_0$ and $h_0$, we find
\begin{equation}
\label{T10} \rho_\mathrm{turb} = t^{-3 \gamma_\mathrm{turb} h_0}
\left(\frac{5C}{6\left(1 - \frac{5}{2} \gamma_\mathrm{turb}  h_0
\right) } \left( t^{1 - \frac{5}{2}  \gamma_\mathrm{turb}  h_0 }
  - t_\mathrm{in}^{1 - \frac{5}{2} \gamma _\mathrm{turb}  h_0 } \right)
+ C_0 a_0^{\frac{5\gamma _\mathrm{turb} H_0}{2}}
\right)^{-\frac{6}{5}} \, .
\end{equation}
If we consider the case of the phantom-like power law expansion,
$a(t)=a_0 \left(t_s - t\right)^{- h_0}$, we find
\begin{eqnarray}
\label{T11}
&& \rho_\mathrm{turb} = \left(t_s - t\right)^{3 \gamma _\mathrm{turb} h_0} \nonumber \\
&& \quad \times \left(\frac{5C}{6\left(1 + \frac{5}{2} \gamma
_\mathrm{turb}  h_0 \right) } \left( \left(t_s - t\right)^{1 +
\frac{5}{2}  \gamma_\mathrm{turb}  h_0 }
 - \left(t_s - t_\mathrm{in} \right)^{1 + \frac{5}{2}  \gamma_\mathrm{turb}  h_0 }
\right)  + C_0 a_0^{\frac{5 \gamma_\mathrm{turb}
H_0}{2}}\right)^{-\frac{6}{5}}\, .
\end{eqnarray}
Especially when $h_0 = - \frac{2}{3 \gamma_\mathrm{turb} }$ with
$w_\mathrm{turb}<-1$, we find $\rho_\mathrm{turb} \propto \left(
t_s - t \right)^{-\frac{6}{5}}$ when $t\to t_s$. In this case,
from (\ref{TR1}), we also obtain $\rho_\mathrm{rad} \propto \left(
t_s - t \right)^{-\frac{6}{5}}$. Then both $\rho_\mathrm{turb}$
and $\rho_\mathrm{rad}$ increase rather rapidly, although the rate
of  increase is smaller than for the energy density of the phantom
dark
  energy where
$\rho_\mathrm{phantom} \propto \left( t_s - t \right)^{-2}$.

Just for simplicity, we now consider the turbulence of the dark energy with $w_\mathrm{turb}=-1$.
Then Eq.~(\ref{53}) gives
\begin{equation}
\label{T12}
\rho_\mathrm{turb} = \left(\frac{6}{5C}\right)^{\frac{6}{5}} \left(t - t_0\right)\, .
\end{equation}
Here $t_0 \equiv t_\mathrm{in} - \frac{6C_0}{5C}$ and we assume
$t_\mathrm{in}>t_0$. We now consider the simple case where the
contribution from the matter (except radiation generated by the
turbulence) can be neglected and the non-turbulent part of the
dark energy has the constant EoS parameter $w_\mathrm{dark}=-1$,
as for the cosmological constant. We write
$\rho_\mathrm{dark}=\Lambda$.

By using (\ref{T12}), we may rewrite (\ref{47}) as
\begin{equation}
\label{T13}
\frac{d}{dt}\left(a^4 \rho_\mathrm{rad} \right)
= \left(\frac{6}{5}\right)^{\frac{11}{5}} C^{-\frac{6}{5}}
\left( t - t_0 \right)^{- \frac{11}{5}} a^4\, ,
\end{equation}
since  $p_\mathrm{rad} = \rho_\mathrm{rad}/3$.
Then by multiplying the FRW equation (\ref{42}) ($\rho_\mathrm{dark}=\Lambda$
and $\rho_\mathrm{matter}=0$) with $a^4$
and differentiating with respect to the cosmological time $t$, we obtain
\begin{equation}
\label{T14}
\frac{6}{\kappa^2}\left(\dot H + 2 H^2\right) = 4\Lambda + 4 \left(\frac{6}{5C}\right)^{\frac{6}{5}}
\left(t- t_0\right)^{-\frac{6}{5}}\, .
\end{equation}
The second term decreases with time and therefore for large $t$, we obtain the asymptotic de Sitter universe, where
$H$ is a constant
\begin{equation}
\label{T15}
H=H_{L0}\equiv \sqrt{\frac{\Lambda\kappa^2}{3}}\, .
\end{equation}
Let assume the turbulence begins at $t=t_0$. Then since we
assume
$w_\mathrm{turb}=w_\mathrm{dark}=-1$, the total dark energy behaves as a cosmological constant and
the de Sitter universe is realized, where
\begin{equation}
\label{T16}
H=H_{I0} \equiv \sqrt{\frac{\kappa^2}{3}\left(\Lambda + \left(\frac{6}{5C}\right)^{\frac{6}{5}}
\left(t_\mathrm{in} - t_0\right)^{-\frac{6}{5}}\right)}\, .
\end{equation}
After the turbulence begins, the cosmological constant decays and $H$
becomes smaller and at the late time,
the universe reaches the asymptotic de Sitter universe with $H=H_{L0}$.
Then we may assume $H_{I0}\gg H_{L0}$.

Thus, the situation  considered is one where there is  matter with
 vanishing EoS parameter and  phantom dark energy. Without
turbulence there could occur a phantom crossing. After such a
crossing, the dark energy dominates.  If there is no non-turbulent
part,  the density of  dark energy is large  and finally satisfies
the condition (\ref{48B}), the dark energy dissipates and converts
into  radiation. Hence, the accelerated expansion will terminate.
Or before satisfying the condition (\ref{48B}), the energy density
goes to a constant (\ref{51}), corresponding to the asymptotic de
Sitter space.

Some remark is in order.
Let us imagine the inflation ended by the turbulence and the turbulent part $\rho_\mathrm{turb}$
of the energy-density generated the reheating. $\rho_\mathrm{turb}$ could
been converted to the radiation. Then the energy-density $\rho_\mathrm{rad}$ of the
radiation after the reheating could be of almost the same order with $\rho_\mathrm{turb}$.
Therefore, we may obtain
\begin{equation}
\label{reheat1}
\rho_\mathrm{rad} \sim \rho_\mathrm{turb}
= \left(\frac{6}{5C}\right)^{\frac{6}{5}} \left(t_\mathrm{in} - t_0\right)\, .
\end{equation}
Here, Eq.~(\ref{T12}) is used.
Applying the Stefan-Boltzmann law
\begin{equation}
\label{reheat2}
\rho_\mathrm{rad} = \sigma T^4\, ,
\end{equation}
with the  constant $\sigma$, one may
evaluate the reheating temperature $T$ as
\begin{equation}
\label{reheat3}
T \sim \sigma^{-\frac{1}{4}}
\left(\frac{6}{5C}\right)^{\frac{3}{10}} \left(t_\mathrm{in} - t_0\right)^{\frac{1}{4}}\, ,
\end{equation}
which may give a constraint on the parameters.
This may indicate on observational manifestations of turbulence.

\section{Discussion}

In summary, we have investigated the role which may be played by
isotropic turbulence in a dark fluid universe at late times. Using
 experimental information about decay of isotropic turbulence in
classical hydrodynamics, two different approaches to modification
of FRW equations are proposed. The consequences of turbulence
presence at future phantom/quintessence universe are studied. It
is demonstrated that it may change the characteristics details of
finite-time future singularity, for instance, the behavior of
scale factor at Big Rip. However, it seems that turbulence cannot
remove the future finite-time singularity in a perturbative
approach. In a non-perturbative approach, the turbulence presence
may change the evolution of dark fluid via its dissipation. In
particular, it may terminate the accelerated evolution or convert
the phantom-dominated universe into a future de Sitter space.
Hence, inclusion of  turbulence may suggest a way to resolve the
future singularity problem.

In principle, one should not limit oneself only to the known decay
law for isotropic turbulence. It is quite possible that at large
scales more general laws should be implemented. In this case, the
description is somewhat similar to the inhomogeneous equation of
state of the universe introduced in
Refs.~\cite{Nojiri:2005sr,Capozziello:2005pa,Nojiri:2006zh}.
Moreover, the explicit scenario of the (effective) turbulence
emergence should be developed. Perhaps, the easiest way to realize
it is to use of cosmological reconstruction in modified gravity
(for a recent review, see \cite{Nojiri:2010wj}). In this case, the
use of turbulence turns out to be an  effective description due to
a corresponding modification of gravity. Alternatively, the origin
of turbulence in a dark fluid universe may be related to an
inhomogeneous equation of state for the universe.

The important question in connection with our proposal of
including turbulence in a dark energy universe is of course
related to the search of observational evidence for turbulence
signatures. This is expected to be possible in the near future
when observational data will give us more accurate information
about the cosmological equation of state.

\section*{Acknowledgments \label{Ack}}

This research has been supported in part
by MEC (Spain) project FIS2006-02842 and AGAUR(Catalonia) 2009SGR-994
(SDO), by Global COE Program of Nagoya University (G07)
provided by the Ministry of Education, Culture, Sports, Science \&
Technology and by the JSPS Grant-in-Aid for Scientific Research (S)
\# 22224003 (SN). The support of the ESF Casimir Network ia also acknowledged.


\end{document}